\documentclass[prl,twocolumn,showpacs,preprintnumbers,amsmath,amssymb]{revtex4}

\usepackage{graphicx}
\usepackage{dcolumn}
\usepackage{bm}

\begin{document}

\title{
``Doubly-magic'' conditions in ``magic''-wavelength trapping of ultracold alkalis
}
\author{Andrei Derevianko}
\email{andrei@unr.edu}
\affiliation{Department of Physics, University of Nevada, Reno NV 89557}

\date{\today}

\begin{abstract}
In experiments with trapped atoms, atomic energy levels
are shifted by the trapping optical and magnetic fields. Regardless of this strong
perturbation, precision spectroscopy
may be still carried out using  specially crafted, ``magic'' trapping fields.
Finding these conditions for particularly valuable microwave clock transitions in alkalis has so far remained an open challenge.
Here I demonstrate that the microwave clock transitions for alkalis may be indeed made  impervious to both
trapping laser intensity and fluctuations of magnetic fields.
I consider driving multiphoton transitions between the clock levels and show that these ``doubly-magic''
conditions are realized at special values of trapping laser wavelengths and fixed values of relatively weak magnetic fields. This finding has implications for precision measurements and quantum information processing with qubits stored in hyperfine manifolds.
\end{abstract}





\maketitle
Compared to spectroscopic beam experiments, trapping cold atoms and molecules removes Doppler shifts and increases interrogation time thereby dramatically enhancing spectral resolution. This improvement comes at a price: trapping optical fields strongly perturb atomic energy levels - transition frequencies are shifted away from their unperturbed values. In addition,
the underlying Stark shift is proportional to the local intensity of the trapping lasers; the shift is non-uniform across the
atomic ensemble and it is also sensitive to laser intensity fluctuations. So trapping seems to be both advantageous and detrimental for precision measurements.
This dilemma is elegantly solved using so-called ``magic'' traps~\cite{YeKimKat08}. At the
``magic'' trapping conditions two levels of interest are shifted by exactly same amount by the trapping fields; therefore the differential effect of trapping fields simply vanishes for that transition.

The idea of such ``magic'' trapping has been crucial for establishing a new class of atomic clocks~\cite{KatTakPal03}, the optical lattice clocks. Here atoms are trapped in optical lattices formed by counter-propagating laser beams; the lasers operate at the magic wavelength.
In these clocks one employs  optical transitions in divalent atoms, such as Sr and Yb.
Finding similar magic conditions for ubiquitous alkali-metal atoms employed in
a majority of cold-atom experiments remains an open challenge. Especially
valuable are the microwave transitions in the ground-state hyperfine manifold. Identifying magic conditions here, for example, would enable
developing microMagic clocks~\cite{BelDerDzu08Clock}: lattice clocks operating in the microwave region of the spectrum.
 In addition, the hyperfine manifolds are used to
store quantum information in a large fraction of quantum computing
proposals with ultracold alkalis. Finding magic conditions would enable a decoherence-free trapping for this important realizations of qubits.

The clock transitions in divalent atoms are between non-magnetic states; this removes sensitivity to magnetic fields.
For alkalis, however, an additional piece of the puzzle  is that the clock/qubit states are sensitive to both optical and magnetic fields.
One needs to eliminate the sensitivity of transition frequency $\nu$ to both perturbations simultaneously. This problem is solved here. We will require that the clock transition is insensitive to both Stark- and Zeeman-induced perturbations (we will use the ``doubly-magic'' qualifier for such trapping conditions).

\begin{figure}[h]
\begin{center}
\includegraphics*[scale=0.65]{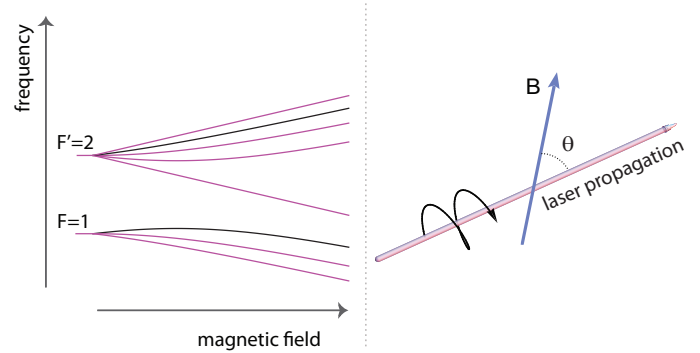}
\end{center}
\caption{(Color online)
Left panel: Zeeman effect for the hyperfine manifold in the ground state of $I=3/2$ isotopes of alkalis. Two clock levels $|F'=2, M'_F=+1 \rangle$ and  $|F=1,M_F=-1\rangle$ are shown in
black. Right panel illustrates geometry of laser-atom interaction: degree of circular polarization, angle $\theta$, and laser wavelength may be varied.
 \label{Fig:GeometryZeeman}
 }
\end{figure}

First steps in identifying magic conditions for hyperfine transitions in
alkalis have been made in Refs.~\cite{RosGheDzu09,FlaDzuDer08,ChoCho07}. That work focused on
eliminating  sensitivity to laser intensity $I_L$ by tuning the laser frequency to its magic value $\omega_m$: $\delta \nu(\omega_m)=0$ regardless of the value of $I_L$ .
The proposals~\cite{FlaDzuDer08,ChoCho07} have neglected the effect
of magnetic fields and focused on B-field sensitive $M_{F}\neq0$ states. So while the trapping
would be Stark-insensitive, the states would decohere due to coupling to stray B-fields.
A partial solution to this problem was discussed in  Refs.~\cite{LunSchPor10,Der10Bmagic}:
moving to the $M_{F}=0$ states eliminates sensitivity to Zeeman shifts to the leading
order. Yet one needs to apply a bias magnetic field of a specific value making the conditions
``magic'' for a given trapping laser wavelength. As a result, the transitions remain Zeeman-sensitive
through the second-order effects; numerical estimates show that, unfortunately, the residual B-field sensitivity would preclude designing a competitive clock.

Clearly, the sensitivity to B-fields has to be addressed. We require that at the magic B-field $d\nu/dB (B_m) = 0$.
Such conditions occur, for example, for a two-photon
$|F'=2, M'_F=+1 \rangle \rightarrow |F=1,M_F=-1\rangle$ transition  in $^{87}$Rb
at the field of about 3 Gauss.
The relevant Breit-Rabi diagram is shown in Fig.~\ref{Fig:GeometryZeeman}.
The two
clock levels are highlighted: the existence of the magic B-field may be inferred visually.
Experimentally, such magic conditions  have been proven  instrumental for performing collisional studies in a
magnetically-trapped Bose-Einstein condensates of $^{87}$Rb~\cite{HarLewMcG02} and studying decoherence of cold cloud of Rb near microchip~\cite{TreHomSte04}. Similar ideas are essential to realizing atomic clocks on a microchip~\cite{LacReiRam10,DeuRamLac10}.
My present work adds the optical fields
to the mix: it turns out that the unified Stark-Zeeman description is non-trivial due to  an interference of the two effects.

{\em Formalism ---}
We are interested in the clock transition between two hyperfine
states $\left\vert F^{\prime}=I+1/2,M_{F}^{\prime}\right\rangle $ and
$\left\vert F=I-1/2,M_{F}\right\rangle$ attached to the ground electronic $nS_{1/2}$ state of an alkali-metal atom ($I$ is the nuclear spin). Here and below we  denote
the upper clock state as $|F'\rangle$ and the lower state as $|F\rangle$ (see Fig.~\ref{Fig:GeometryZeeman}).

We focus on the $M_{F}'=-M_{F}$ transitions. For these transitions,
the electronic $g$-factors of the
two states are the same (see Fig.~\ref{Fig:GeometryZeeman}). Then the bulk of the Zeeman shift of the transition frequency goes away and the linear Zeeman effect is determined only by the nuclear $g$-factor $g_I = 1/I \,\mu_{\rm nuc}/\mu_N$, where $\mu_N$ is the nuclear magneton. The residual linear shift is compensated by the second-order Zeeman correction, quadratic in the B-field. This leads to the magic value of
 the B-field
\begin{equation}
 B_{m}\approx
\frac{g_{I}\mu _{N}~M_{F^{\prime }}}{2\left\vert \langle F,M_{F^{\prime }}\left\vert \mu _{z}^{e}\right\vert F^{\prime },M_{F^{\prime }}\rangle \right\vert ^{2}} \, h\nu_\mathrm{clock} \,,
\label{Eq:Bmagic}
\end{equation}
where $\mu^e$ is the operator of magnetic moment of electron.
The second-order estimate, Eq.(\ref{Eq:Bmagic}), is a good approximation:
for $^{87}$Rb it gives 3.25 G, while the ``all-order'' Breit-Rabi analysis yields $B_{m} = 3.228917(3) \, \mathrm{G}$, Ref.~\cite{HarLewMcG02}.
Values of $B_m$ for Rb and Cs isotopes are tabulated in Table~\ref{Tab:BmList}.
Notice that the fields for all tabulated transitions are relatively weak and can be well
stabilized using existing technologies~\cite{LacReiRam10}.

\begin{table}[h]
\caption{ Values of ``magic'' B-fields and ``magic'' wavelengths.
 \label{Tab:BmList}}
\begin{ruledtabular}
\begin{tabular}{ccc}
Transition  &
$B_m$, Gauss & $\lambda_m$ \\
\hline
\multicolumn{3}{c}{
$^{87}$Rb,  $I=3/2$,  $\nu_\mathrm{clock}=6.83 \, \mathrm{GHz}$} \\
$|2,1\rangle \to  |1,-1\rangle$ & 3.25  & 806 nm\footnotemark[1]  \\[1ex]
\multicolumn{3}{c}{
$^{85}$Rb,  $I=5/2$,  $\nu_\mathrm{clock}=3.04 \, \mathrm{GHz}$} \\
$|3,1\rangle \to  |2,-1\rangle$ &  0.359 &  --- \\
$|3,2\rangle \to  |2,-2\rangle$ &  1.15  & 479--658;797--878 \\[1ex]
\multicolumn{3}{c}{
$^{133}$Cs,  $I=7/2$,  $\nu_\mathrm{clock}=9.19 \, \mathrm{GHz}$} \\
$|4,1\rangle \to  |3,-1\rangle$ & 1.41   &  --- \\
$|4,2\rangle \to  |3,-2\rangle$ & 3.51  & 906--1067; 560-677\\
$|4,3\rangle \to  |3,-3\rangle$ & 9.04  & 898--1591;863--880; 512--796\\
\end{tabular}
\end{ruledtabular}
\footnotetext[1]{nearly doubly-magic}
\end{table}

Fixing magnetic field at its magic value accomplishes the Zeeman-insensitivity of the clock transitions.
Now we would like to additionally remove the Stark sensitivity to  intensity of trapping
laser fields.  We consider the following setup shown in Fig.~\ref{Fig:GeometryZeeman}.
An atom is illuminated by a laser light, with a certain degree of circular polarization $A$.
At the same time, a bias magnetic field is applied at an angle $\theta$ to the direction
of laser propagation. The B-field is fixed at its magic value. This is a basic building block for optical trapping. For example.
an optical lattice (standing wave) may be formed by two counter-propagating lasers of the same wavelength.

In a laser field, both clock levels are shifted due to the dynamic (i.e.,
laser-frequency-dependent) Stark effect~\cite{ManOvsRap86}.
I derived the following formula for the differential shift of the clock
frequency\footnote{The derivation is based on formalism developed in Ref.~\cite{Der10Bmagic}},
\begin{widetext}
\begin{equation}
\delta \nu_{\mathrm{clock}} (\omega)=-
 \frac{1}{h} \left\{
\left(  \beta_{F^{\prime}}^{s}-\beta_{F}^{s}\right)  +
  A\cos\theta~M_{F^{\prime}} \left[  \left(  \frac
{1}{2F^{\prime}}\beta_{F^{\prime}}^{a}+\frac{1}{2F}\beta_{F}^{a}\right)
+
g_{I}\frac{\mu_{N}}{\mu_{B}}~\alpha_{nS_{1/2}}^{a} \right]  \right\}  \left(  \frac{E_{L}}{2}\right)  ^{2} \, . \label{Eq:ClockShiftMultiPhoton}
\end{equation}
\end{widetext}
Here $E_L$ is the amplitude of laser E-field, $I_L \propto E_L^2$. The quantities $\beta^s$, $\beta^a$, and $\alpha^a$ are the
scalar and vector (axial) polarizabilities (see below). An important fact is that all these polarizabilities depend on the laser frequency, $\omega$.
By tuning the laser frequency we require that the combination in curly bracket becomes zero.
At that ``magic'' point, the differential clock shift vanishes independently of the laser intensity: $\delta \nu_{\mathrm{clock}} (\omega_m) =0$.

What is the difference between the polarizabilities $\alpha$ and $\beta$?
We are considering the Stark shift of hyperfine levels attached to the same electronic
state. To the leading order, the shift is determined by the properties of the underlying electronic state (polarizability $\alpha$). However, because the electronic state for
both hyperfine levels is the same, the levels are shifted at the same rate\footnote{This statement holds for scalar polarizabilities and also, because of opposite g-factors, in the case of $M_F=-M_F'$ for vector
polarizabilities. Tensor polarizability vanishes for the $nS_{1/2}$ states.}
and we need
to  distinguish between the two hyperfine levels. An apparent difference between the two
clock levels is caused by the hyperfine interaction (HFI), and the rigorous analysis involves
so-called HFI-mediated polarizabilities, $\beta$. Lengthy third-order (two dipole couplings
to the laser field and one HFI) expressions
for these polarizabilities may be found in Ref.~\cite{RosGheDzu09}.

Continuing with our discussion of the clock shift~(\ref{Eq:ClockShiftMultiPhoton}), I would like to stress an unconventional origin of the last contribution. Its full form for an arbitrary B-field is
$A\cos\theta~M_{F^{\prime}} g_{I}\frac{\mu_{N}}{\mu_{B}}~\alpha_{nS_{1/2}}^{a} \times
\left( B/B_m\right)$. The term arises due to an interference between Stark and Zeeman interactions. Qualitatively, the vector contribution to the Stark shift has the very same rotational
properties as the Zeeman coupling (both are vector operators). These operators, in particular,
couple the two hyperfine manifolds. Consider shift of the $|F',M_F\rangle$ level. The Zeeman operator couples it to the $|F,M_F\rangle$ intermediate state, and then the vector Stark shift operator
brings it back to the $|F',M_F\rangle$ level thereby resulting in the energy shift. This cross-term is of the same order of magnitude
as the other two terms in Eq.~(\ref{Eq:ClockShiftMultiPhoton}) and has to be included in the consideration.

Now we can find magic wavelengths by numerically evaluating
atomic polarizabilities entering Eq.~(\ref{Eq:ClockShiftMultiPhoton}).
To this end I used a blend of relativistic many-body
techniques of atomic structure, as described in~\cite{BelSafDer06}.
To improve upon the accuracy, high-precision experimental data
were used where available. To ensure the quality of the calculations, a
comparison with the experimental literature data on static Stark shifts of the
clock transitions was made.
It is one of the problems where a consistent treatment is important and less sophisticated estimates may fail even qualitatively, see~\cite{Der10Bmagic} for
a discussion.

{\em Results --- }
I will present the results of the calculations in the following form. Since the magic
condition corresponds to $\delta \nu_{\mathrm{clock}} (\omega_m) =0$, we may recast
Eq.~(\ref{Eq:ClockShiftMultiPhoton}) into
\begin{equation}
M_{F^{\prime }}A\cos \theta =-\frac{\beta _{F^{\prime }}^{s}-\beta _{F}^{s}}{\left( \frac{1}{2F^{\prime }}\beta _{F^{\prime }}^{a}+\frac{1}{2F}\beta _{F}^{a}\right) +
g_{I}\frac{\mu _{N}}{\mu _{B}}\alpha _{nS_{1/2}}^{a}} \, .
\label{Eq:MFAcos}
\end{equation}
The r.h.s.\ of this equation depends on the laser frequency, while the l.h.s.\ does not.
Moreover, $|A\cos \theta| \le 1$, therefore the magic conditions would exist
only if for a given $\omega$  the r.h.s.\ is within the range $-|M_F'|$ and $|M_F'|$.

\begin{figure}[h]
\begin{center}
\includegraphics*[scale=1.0]{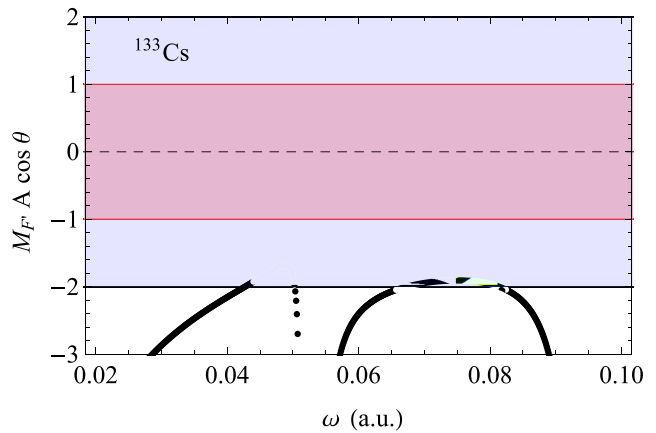}
\end{center}
\caption{(Color online) Magic conditions for $^{133}$Cs. A dependence of the product
$M_{F^{\prime }}A\cos \theta$ on trapping laser frequency (in atomic units) is plotted. The shaded regions
are bound by $-|M_{F^{\prime }}|$ and $+|M_{F^{\prime }}|$ lines. Magic trapping for
a $|F'=4,M_F'\rangle \to |F=3,-M_F'\rangle$ clock transition is only possible when
the computed curve lies inside the corresponding shaded region.
 \label{Fig:133CsAcos}
 }
\end{figure}

Doubly-magic trapping of $^{133}$Cs is analyzed in Fig.~\ref{Fig:133CsAcos}. This atom is metrologically important.
For the past four decades the SI unit of time, the
second, has been defined as a duration of a certain number of periods of radiation
corresponding to the transition between the two hyperfine levels of the ground state of the $^{133}$Cs atom.
Cs clocks serve as primary frequency standards worldwide
and there is a substantial investment in the infrastructure supporting these clocks.
The most accurate Cs clocks are fridge-sized fountain clocks (see, e.g., Ref.~\cite{WynWey05}).
Developing microwave lattice clocks based on Cs may be beneficial as the active chamber
of the clock will be reduced to a few micrometers across. This million-fold reduction in size
is anticipated to lead to a better control over detrimental black body radiation and stray magnetic fields.

From Fig.~\ref{Fig:133CsAcos} we find that the doubly-magic trapping of Cs atoms
is indeed possible for two transitions: $|4,2\rangle \to  |3,-2\rangle$ and
$|4,3\rangle \to  |3,-3\rangle$. The only complication is that driving the former transition requires 4 photons,
while the latter transition requires 6 photons. This may be potentially
accomplished either with multi-step RF/MW or stimulated Raman drives~\cite{HarLewMcG02,AlePaz97}.

The curve (\ref{Eq:MFAcos}) exhibits a resonant behavior when the laser frequency
passes through the fine-structure doublet of atomic transitions: $6s_{1/2}-6p_{1/2}$ and $6s_{1/2}-6p_{3/2}$
at $\omega=0.050932$ a.u. and $0.053456$ a.u.. Not shown in the
Fig.~\ref{Fig:133CsAcos} are the values of the r.h.s.\ of Eq. (\ref{Eq:MFAcos}) between
the resonances. In this region, the r.h.s.\ values
become positive and the curve crosses the $M_{F^{\prime }}A\cos \theta=+3$ limit from above:
the 6-photon transition may be made doubly-magic when the laser is tuned to inside the fine-structure doublet.

\begin{figure}[h]
\begin{center}
\includegraphics*[scale=1.0]{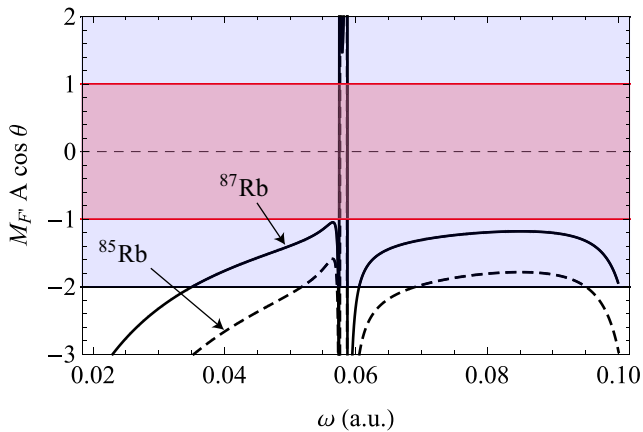}
\end{center}
\caption{(Color online) Same as in Fig.~\ref{Fig:133CsAcos} for $^{87}$Rb transition  $|F'=2,1\rangle
\to |F=1,-1\rangle$ and for $^{85}$Rb transitions  $|F'=3,1\rangle \to |F=2,-1\rangle$ and $|F'=3,2\rangle \to |F=2,-2\rangle$.
 \label{Fig:Rb8785Acos}
 }
\end{figure}

$^{87}$Rb serves as the secondary frequency standard. An analysis of magic conditions
for this atom is carried out in Fig.~\ref{Fig:Rb8785Acos}. There is a single, $|F'=2,1\rangle \to |F=1,-1\rangle$, transition of interest here. This is a two-photon transition.
Curiously, the  $M_{F^{\prime }}A\cos \theta$ curve nearly touches its limiting value at $\omega \approx 0.0565$ a.u.
($\lambda_m \approx 806\, \mathrm{nm})$ somewhat below the $5s_{1/2}-5p_{3/2}$ resonance. Here the r.h.s.\ of Eq.~(\ref{Eq:MFAcos}) reaches values of
$\approx -1.05$, i.e., it is just 5\% off the limiting value of -1. While not
quite achieving the ``doubly-magic'' status, this 806 nm wavelength gets us to nearly-magic conditions. In practice, if one may afford small
Zeeman decoherences, the bias B-field may be detuned off its magic value. Then (see discussion above) the last term in the Stark shift, Eq.(\ref{Eq:ClockShiftMultiPhoton}), is rescaled by the ratio $(B/B_m)$ and
the magic condition for the Stark shift can be reached. For example, I computed that
$\lambda_m \approx 806$ nm becomes ``Stark-magic'' at B-field of 3.45 G, i.e.,
just  6\% larger than its ``Zeeman-magic'' value. Ultimately, the choice of $B$ should be a matter of optimizing tolerances to both Stark and Zeeman-induced decoherence in a particular application.
For example, in a recent Paris experiment~\cite{DeuRamLac10} with a magnetically trapped
ensemble of $^{87}$Rb atoms, the bias B-field has been varied by as much as 16\% from its magic value; this still has led to well-resolved contrast on the  $|F'=2,1\rangle \to |F=1,-1\rangle$ clock transition.

Finally, in the same Fig.~\ref{Fig:Rb8785Acos} we explore magic conditions for another isotope of Rb, $^{85}$Rb. Compared to $^{87}$Rb, the nuclear spin of this isotope is larger ($I=5/2$), this
results in larger values of $F$ and richer magnetic substructure of the hyperfine levels.
Two transitions of interest become available: $|F'=3,1\rangle \to |F=2,-1\rangle$ and $|F'=3,2\rangle \to |F=2,-2\rangle$.
Moreover, various contributions to Eq.~(\ref{Eq:ClockShiftMultiPhoton}) scale differently with the nuclear spin and we need to carry out separate calculation for each isotope.
From Fig.~\ref{Fig:Rb8785Acos}, we find that doubly-magic conditions can be attained for the $|F'=3,2\rangle \to |F=2,-2\rangle$ transition in $^{85}$Rb.

I also carried out calculations for other commonly-used alkalis. For $^7$Li, $^{23}$Na,
 and $^{39}$K there are no doubly-magic (or ``near-magic'') points; all these
  isotopes have $I=3/2$. For various Fr isotopes,
 there is a multitude of doubly-magic points for multi-photon transitions. For example,
 for $^{210}$Fr ($I=6$) transition $M_F = -7/2 \to +7/2$  these conditions
 are attained in the range $\lambda_m =846-1061$ nm.

To conclude, by working at the doubly-magic (Stark-Zeeman) conditions, one can greatly reduce sensitivity to spatial inhomogeneity due to trapping/bias fields and also reduce sensitivity
to temporal fluctuations of the fields.
It is anticipated that a variety of applications could take advantage of the magic (and nearly-magic) conditions found in this paper. For example, we anticipate that lifetimes of quantum memory~\cite{ZhaDudJen09} may be improved. Another opportunity is developing microMagic lattice clocks. The important finding here is that multiphoton transitions in
metrologically-important $^{133}$Cs can be made simultaneously insensitive to both intensities of trapping lasers and also to fluctuations of magnetic fields.

\emph{Acknowledgements ---}
 I would like to thank  Peter Rosenbusch, Kurt Gibble, Trey Porto, and Hartmut H\"{a}ffner for  discussions.
 This work was supported in part by the NSF and by the NASA under Grant/Cooperative Agreement No. NNX07AT65A issued by the Nevada NASA EPSCoR program.


\end{document}